\documentclass[twocolumn,showpacs,preprintnumbers,amsmath,amssymb,
superscriptaddress, groupedaddress]{revtex4}

\usepackage{graphicx}
\usepackage{dcolumn}
\usepackage{bm}

\begin{document}

\title{Universal spreading of wavepackets in disordered nonlinear systems}

\author{S. Flach}
\affiliation{Max Planck Institute for the Physics of Complex Systems, 
N\"othnitzer Str. 38, D-01187 Dresden, Germany}
\author{D. O. Krimer}
\affiliation{Max Planck Institute for the Physics of Complex Systems, 
N\"othnitzer Str. 38, D-01187 Dresden, Germany}
\author{Ch. Skokos}
\affiliation{Max Planck Institute for the Physics of Complex Systems, 
N\"othnitzer Str. 38, D-01187 Dresden, Germany}

\date{\today}

\begin{abstract}
In the absence of nonlinearity all eigenmodes of a chain with disorder 
are spatially localized
(Anderson localization).
The width of the eigenvalue spectrum, and the average eigenvalue spacing
inside the localization volume, set two frequency scales.
An initially localized wavepacket spreads in the presence of nonlinearity.
Nonlinearity introduces frequency shifts, which define three
different evolution outcomes: i) localization as a transient, with subsequent subdiffusion;
ii) the absence of the transient, and immediate subdiffusion; iii) selftrapping of a part of the packet,
and subdiffusion of the remainder.
The subdiffusive spreading is due to a finite number of packet modes being resonant.
This number does not change on average, and depends only on the disorder strength.
Spreading is due to corresponding weak chaos inside the packet, which slowly heats the
cold exterior. 
The second moment of the packet is increasing as $t^{\alpha}$. We find $\alpha=1/3$.
\end{abstract}

\maketitle
The normal modes (NM) of a one-dimensional linear system
with uncorrelated random potential are spatially localized (Anderson localization).   
Therefore any wavepacket, which is initially localized,
remains localized for all time \cite{PWA58}.
When nonlinearities are added, NMs interact with each other \cite{GrKiv92}. 
Recently, experiments were performed on light
propagation in spatially random nonlinear optical media \cite{Exp} and
on Bose-Einstein condensate expansions in random optical potentials \cite{BECEXP}.

Numerical studies of wavepacket propagation in several models 
showed that the second moment of the norm/energy 
distribution grows subdiffusively in time as $t^{\alpha}$ \cite{Shep93Mol98,PS08},
with $\alpha \approx 1/3$. Reports on partial localization
were published as well \cite{kkfa08},\cite{shapiro}. 
Further numerical conductivity studies report Ohmic behaviour at finite energy densities
\cite{abjl08}.

The aim of the present work is to clarify the mechanisms of wavepacket 
spreading and localization.
We study two models.  
The Hamiltonian of the disordered discrete nonlinear Schr\"odinger equation (DNLS) 
\begin{equation}
\mathcal{H}_{D}= \sum_l \epsilon_l |\psi_l|^2+\frac{\beta}{2} |\psi_l|^{4}
-  (\psi_{l+1}\psi_{l}^{\star} +c.c.)
\label{RDNLS}
\end{equation}
with complex variables $\psi_l$. 
The random on-site energies $\epsilon_l$ are  
chosen uniformly from the interval $\left[-\frac{W}{2},\frac{W}{2}\right]$.  
The equations of motion are generated by $\dot{\psi}_l = \partial \mathcal{H}_{D}/
\partial (i \psi^{\star}_l)$.

The Hamiltonian of the quartic Klein-Gordon chain (KG) of coupled anharmonic
oscillators with coordinates $u_l$ and momenta $p_l$
\begin{equation}
\mathcal{H}_{K}= \sum_l  \frac{p_l^2}{2} +\frac{\tilde{\epsilon}_l}{2} u_l^2 + 
\frac{1}{4} u_l^4+\frac{1}{2W}(u_{l+1}-u_l)^2\;.
\label{RQKG}
\end{equation}
The equations of motion are $\ddot{u}_l = - \partial \mathcal{H}_{K} /\partial u_l$,
and $\tilde{\epsilon}_l$
are chosen uniformly from the interval $\left[\frac{1}{2},\frac{3}{2}\right]$.

We consider a wavepacket at $t=0$ which is given by a single site excitation
$\Psi_l = \delta_{l,l_0}$ with $\epsilon_{l_0}=0$ for DNLS, and 
$x_l=X\delta_{l,l_0}$ with $p_l=0$ and $\tilde{\epsilon}_{l_0}=1$ for KG.
The value of $X$ controls the energy $E$ in the latter case. 

We will use the DNLS for theoretical considerations, and present numerical results for both models \cite{equiv}.
For $\beta=0$ (\ref{RDNLS}) is reduced to the linear eigenvalue problem
$\lambda A_l = \epsilon_l A_l -(A_{l+1} + A_{l-1})$. The eigenvectors $A_{\nu,l}$
are the NMs, and the eigenvalues $\lambda_{\nu}$ are the frequencies of the NMs.

The width of the spectrum $\{ \lambda_{\nu} \}$ is $\Delta=W+4$.
The asymptotic spatial decay of an eigenvector is given by $A_{\nu,l} \sim {\rm e}^{-l/\xi }$
where $\xi(\lambda_{\nu}) \leq \xi(0) \approx 100/W^2$ is the localization length
\cite{KRAMER}. The NM participation number $P_{\nu} = 1/\sum_l A_{\nu,l}^4$  
characterizes the spatial extend - localization volume - of the NM. It is distributed around
the mean value $\overline{P_{\nu}}\approx 3.6 \xi$ with variance $\approx (1.3 \xi)^2$ \cite{MIRLIN}.
The average spacing 
of eigenvalues of NMs within the range of a localization volume is therefore
$\overline{\Delta \lambda} \approx \Delta / \overline{P_{\nu}} $.
The two scales $ \overline{\Delta \lambda} < \Delta $ determine the packet evolution details in the presence
of nonlinearity.
  
The equations of motion of (\ref{RDNLS})  in normal mode space read
\begin{equation}
i \dot{\phi}_{\nu} = \lambda_{\nu} \phi_{\nu} + \beta \sum_{\nu_1,\nu_2,\nu_3}
I_{\nu,\nu_1,\nu_2,\nu_3} \phi_{\nu_1} \phi_{\nu_2}^* \phi_{\nu_3}\;
\label{NMeq}
\end{equation}
with the overlap integral 
\begin{equation}
I_{\nu,\nu_1,\nu_2,\nu_3} = \sum_l A_{\nu,l} A_{\nu_1,l} A_{\nu_2,l} A_{\nu_3,l}\;.
\label{OVERLAP}
\end{equation}
The variables $\phi_{\nu}$ determine the complex time-dependent amplitudes of the NMs.  

The nonlinear frequency shift
at site $l_0$ is $\delta \lambda \approx \beta$.
Then we expect three qualitatively different regimes of spreading:
i) $\beta < \overline{\Delta \lambda}$; ii) $\overline{\Delta \lambda} < \beta < \Delta$;
iii) $\Delta < \beta$. In case i) the local frequency shift is less than the average spacing
between excited modes, therefore no initial resonance overlap of them is expected,
and the dynamics may - at least for long times - evolve as the one for $\beta=0$.
In case ii) resonance overlap may happen immediately, and the packet should evolve differently. 
For iii) the frequency shift exceeds the spectrum width, therefore 
some renormalized frequencies of NMs (or sites) 
may be tuned out of resonance with the NM spectrum, leading to selftrapping.
The above definitions are highly qualitative, since localized initial conditions are subject
to strong fluctuations. 
Yet, regime iii) is also captured by a theorem presented in \cite{kkfa08}, which proves,
that for $\beta > \Delta$ the single site excitation can not uniformly spread over the entire
(infinite) lattice for the DNLS case. 

We order the NMs in space by increasing value of the center-of-norm coordinate 
$X_{\nu}=\sum_l l A_{\nu,l}^2$.
We analyze normalized distributions $z_{\nu} \geq 0$ using the second moment
$m_2= \sum_{\nu} (\nu-\bar{\nu})^2 z_{\nu}$ and the participation number 
$P=1 / \sum_{\nu} z_{\nu}^2$,
which measures the number of the strongest excited sites in $z_{\nu}$.
Here $\bar{{\nu}} = \sum_{\nu} \nu z_{\nu}$.
For DNLS we follow norm density distributions $z_{\nu}\equiv |\phi_{\nu}|^2/\sum_{\nu} |\phi_{\nu}|^2$.
For KG we follow normalized energy density distributions 
$z_{\nu}\equiv h_{\nu}/\sum_{\nu} h_{\nu}$ with $h_{\nu} = \dot{a}^2_{\nu}/2+\omega^2_{\nu}a_{\nu}^2/2$,
where $a_{\nu}$ is the amplitude of the $\nu$th NM and $\omega^2_\nu=1+(\lambda_{\nu}+2)/W$.

\begin{figure}
\includegraphics[angle=0,width=0.99\columnwidth]{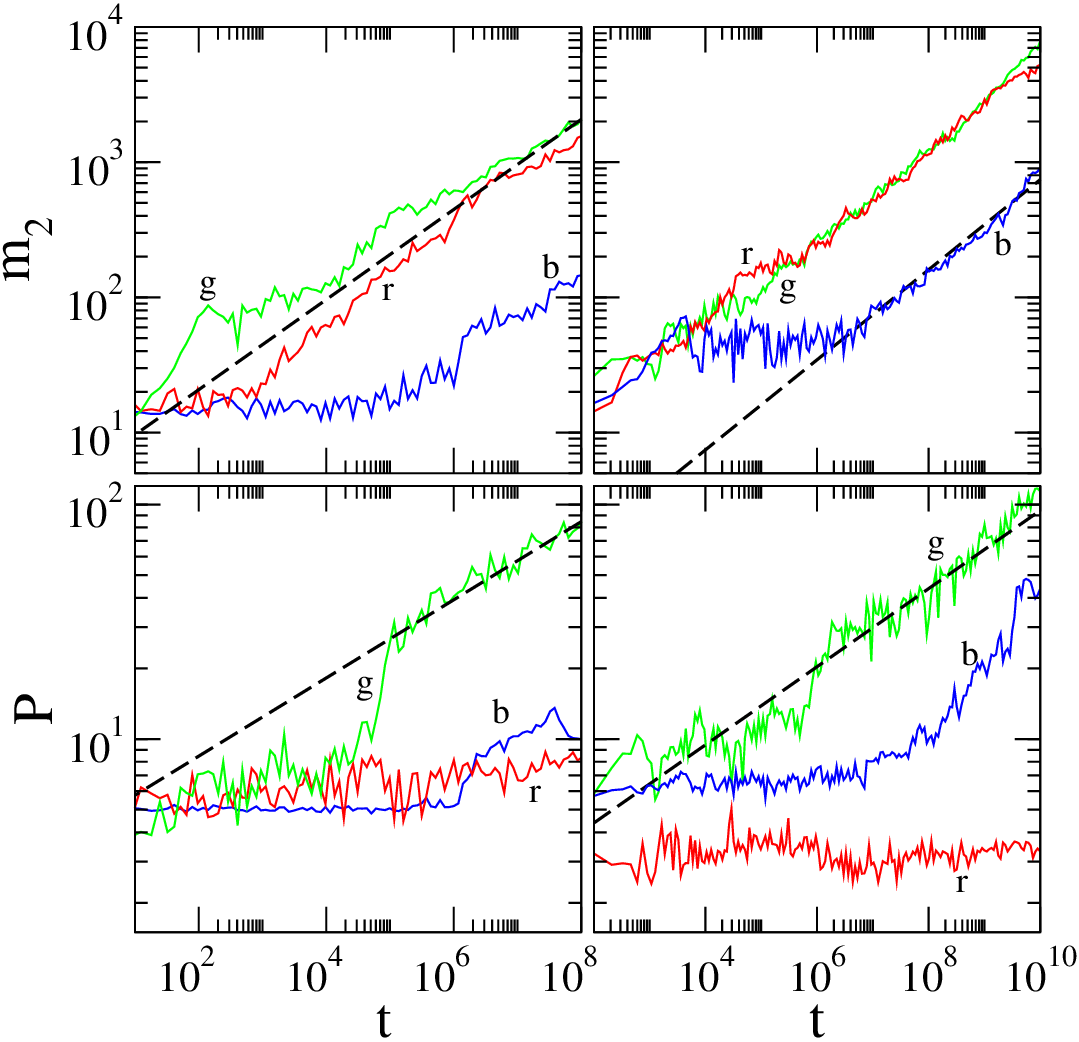}
\caption{(color online) 
$m_2$ and $P$ versus time in log-log plots.
Left plots: DNLS with $W=4$, $\beta=0.1,1,4.5$ ((b)lue, (g)reen, (r)ed). 
Right plots: KG with $W=4$ and initial energy $E=0.05,0.4,1.5$ 
((b)lue, (g)reen, (r)ed).
The disorder realization is kept unchanged for each of the models.
Dashed straight lines guide the eye for exponents 1/3 ($m_2$) and 1/6 ($P$) 
respectively.
}
\label{fig1}
\end{figure}

We systematically studied the evolution of wavepackets for (\ref{RDNLS}) and (\ref{RQKG}) \cite{numerics}.
The above scenario was observed very clearly. Examples are shown in Fig.\ref{fig1}.
Regime iii) yields selftrapping (see also Figs.1,3 in \cite{kkfa08}), therefore
$P$ does not grow significantly, while the second moment 
$m_2\sim t^{\alpha}$ with $\alpha \approx 1/3$ (red curves).
Thus a part of the excitation stays highly localized \cite{kkfa08}, while another part
delocalizes. Note that for large $\beta \gg \Delta$ (or similar energy for the KG model) 
almost all the excitation is selftrapped. 
Regime ii) yields subdiffusive spreading with $m_2\sim t^{\alpha}$
and $P \sim t^{\alpha/2}$ \cite{Shep93Mol98} (green curves).
Regime i) shows Anderson localization up to some time scale $\tau$ which 
increases with decreasing $\beta$. For $t < \tau$ both $m_2$ and $P$ are not changing.
However for $t > \tau$ a detrapping takes place, and the packet starts to grow
with characteristics as in ii) (blue curves). Therefore regime i) is a transient, which ends
at some time $\tau$, and after that regime ii) takes over.

Partial nonlinear localization in regime iii) is explained by
selftrapping \cite{kkfa08}. It is due to tuning frequencies of excitations out of 
resonance with the NM spectrum, takes place irrespective of the presence of
disorder and is related to the presence of exact $t$-periodic 
spatially localized states (also coined discrete breathers) for ordered \cite{DB}
and disordered systems \cite{DDB} (in the latter case also $t$-quasiperiodic states
exist). These exact solutions act as trapping centers.

Anderson localization on finite times in regime i) is 
observed
on potentially large time scales $\tau$, and as in iii), regular
states act as trapping centers \cite{DDB}. For $ t > \tau$, the wavepacket trajectory
finally departs away from the vicinity of regular orbits, and deterministic chaos sets in inside
the localization volume, with subsequent spreading.

\begin{figure}
\includegraphics[angle=0,width=0.99\columnwidth]{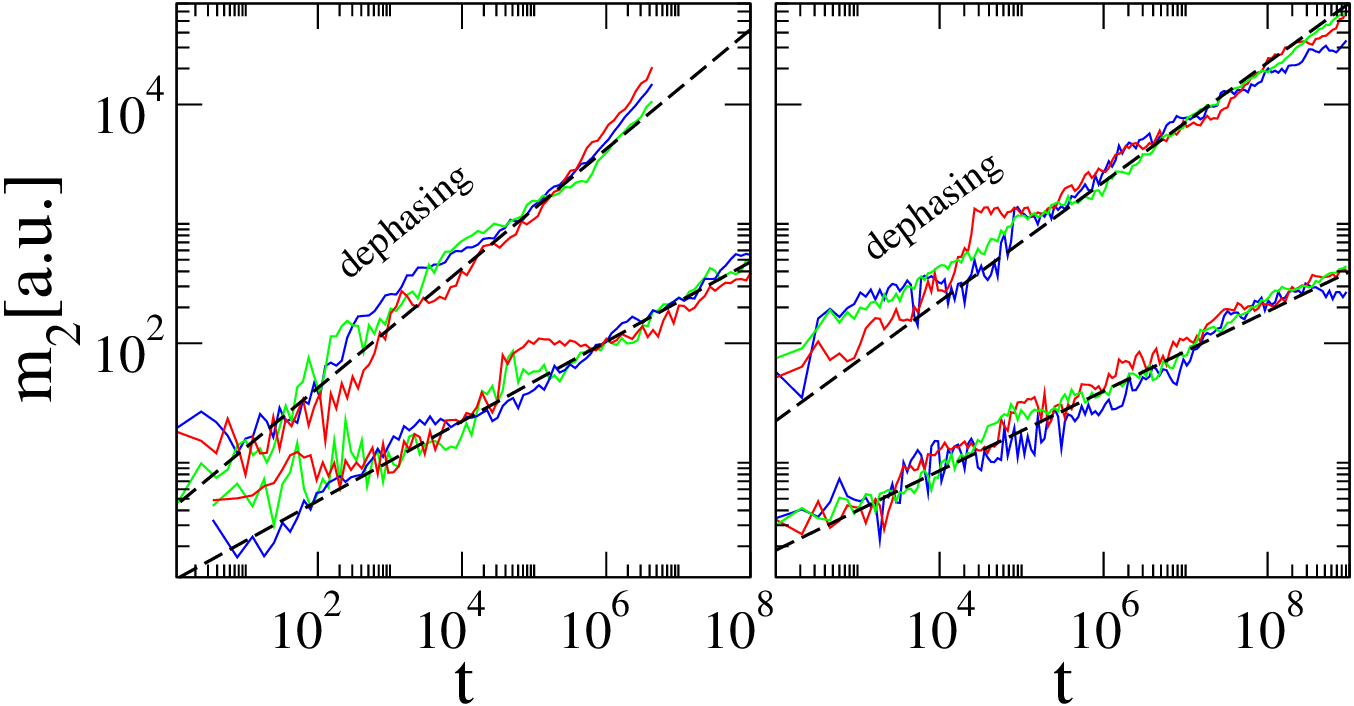}
\caption{(color online) 
$m_2$ (in arbitrary units) versus time in log-log plots in regime ii) and different disorder strength.
Lower set of curves: plain integration (without dephasing); upper set of curves: integration with dephasing
of NMs, see text.
Dashed straight lines with exponents 1/3 (no dephasing) and 1/2 (dephasing) guide the eye. 
Left panel: DNLS,  $W=4$, $\beta=3$(blue); $W=7$, $\beta=4$ (green); $W=10$, $\beta=6$  (red).
Right panel: KG,  $W=10$, $E=0.25$ (blue) , $W=7$, $E=0.3$ (red) , $W=4$, $E=0.4$ (green).
The curves are shifted vertically in order to give maximum overlap
within each group.
}
\label{fig2}
\end{figure}

The subdiffusive spreading takes place in regime i) for $t>\tau$, in regime ii), and for a part
of the wavepacket also in regime iii). 
The exponent $\alpha$ does not appear to depend on $\beta$. In Fig.\ref{fig2}
we show results for $m_2(t)$ in the respective regime ii) for different disorder strength.
Again we find no visible dependence of the exponent $\alpha$ on $W$.
Therefore the subdiffusive spreading is rather universal. 

Let each NM in the packet after some spreading to have norm $|\phi_{\nu}|^2 \sim n \ll 1$.
The packet size is then $1/n \gg \overline{P_{\nu}}$, and the second moment $m_2 \sim 1/n^2$.
We can think of two possible mechanisms of wavepacket spreading. A NM with index $\mu$ in a layer
of width $\overline{P_{\nu}}$ in the cold exterior,
which borders the packet,
is either {\sl heated} by the packet, or {\sl resonantly excited} by some particular NM from a
layer with width $\overline{P_{\nu}}$ inside the
packet. Heating here implies a (sub)diffusive spreading of energy. Note that 
the numerical results yield subdiffusion, supporting the nonballistic diffusive heating
mechanism.

For heating to work, the packet modes should have a continuous frequency part
in their temporal spectrum
(similar to a white noise), at variance to a pure point spectrum.
Therefore at least some NMs of the packet should evolve chaotically in time. 
Ref. \cite{PS08} assumed that all NMs in the packet are chaotic,
and their phases can be assumed to be random. 
With (\ref{NMeq}) the heating of the exterior mode should evolve as 
$i \dot{\phi}_{\mu} \approx \lambda_{\mu} \phi_{\mu} + \beta n^{3/2} f(t)$ where
$\langle f(t) f(t') \rangle = \delta(t-t')$ ensures that $f(t)$ has a continuous frequency spectrum. 
Then the exterior NM is increasing its
norm according to $|\phi_{\mu}|^2 \sim \beta^2 n^3 t$. The momentary diffusion
rate of the packet is given by the inverse time it needs to heat the exterior mode up to the
packet level: $D = 1/T \sim \beta^2 n^2$. The diffusion equation $m_2 \sim D t$ yields
$m_2 \sim \beta t^{1/2}$ \cite{comment1}. We tested the above conclusions by enforcing
decoherence of NM phases \cite{comment2} and obtain $m_2\sim t^{1/2}$
(see Fig.\ref{fig2}).
Therefore, when the NMs dephase completely, the exponent $\tilde{\alpha}=1/2$, {\sl contradicting} 
numerical observations {\sl without dephasing}.
Thus, not all NMs in the packet are chaotic, and dephasing is at best a partial outcome.

\begin{figure}
\includegraphics[angle=0,width=0.99\columnwidth]{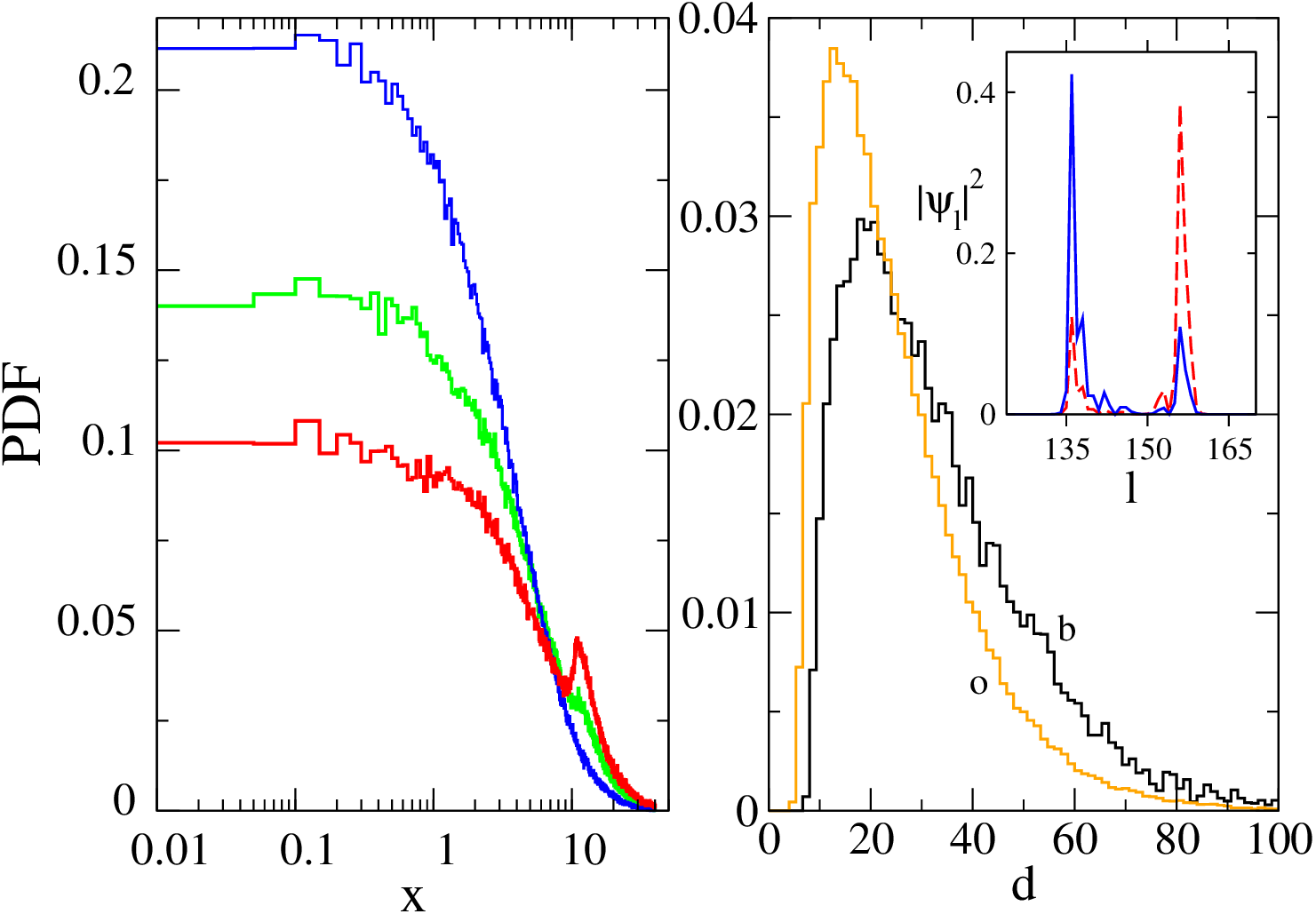}
\caption{(color online) 
Left plot:  Probability densities $\mathcal{W}(x)$ of NMs being resonant 
(see text for details). $W=4,7,10$ (from top to bottom).
Right plot: Probability densities $\mathcal{P}(d)$ of peak distances between
resonant NM pairs for $W=4$; (o)range: $x_c=0.1$, (b)lack: $x_c=0.01$ (see text for details).
Inset: The eigenvectors of a resonant pair of double-peaked states (solid and dashed
lines).
}
\label{fig3}
\end{figure}

Chaos is a combined result of resonances and nonintegrability. Let us
estimate the number of resonant modes in the packet. 
A NM with $|\phi_{\nu}|^2 = n$ will excite other modes in first order in $\beta n$ as
\begin{equation}
|\phi_{\mu}| = \beta n R_{\nu,\mu}^{-1} |\phi_{\nu}|\;,\;
R_{\nu,\mu} \sim \left|\frac{\lambda_{\nu}-\lambda_{\mu}}{I_{\mu,\nu,\nu,\nu}}\right|
\;.
\label{PERT1}
\end{equation}
The perturbation approach breaks down, and resonances set in, when $R_{\nu,\mu} < \beta n$.
We perform a statistical numerical analysis. For a given NM $\nu$ we obtain 
$ R_{\nu,\mu_0} = \min_{\mu \neq \nu} R_{\nu,\mu}$. 
Collecting $R_{\nu,\mu_0}$ for many $\nu$ and many disorder realizations,
we find the probability density distribution $\mathcal{W}(R_{\nu,\mu_0})$ (Fig.\ref{fig3}, left plot).
The main result is that $\mathcal{W}(x \rightarrow 0) \rightarrow const \neq 0$.
Therefore the probability for a mode in the packet to be resonant is proportional to $\beta n$.
On average the number of resonant modes in the packet is constant, proportional to $\beta$,
and their fraction within the packet is $\sim \beta n$. 
Since packet mode amplitudes fluctuate in general, averaging is meant
both over the packet, and over suitably long time windows (yet short compared
to the momentary inverse packet growth rate).
We conclude,
that the continuous frequency part of the dynamics of a packet mode is scaled down by $\beta n$,
compared to the case when all NMs would be chaotic. 
Then the exterior NM is heated according to
$i \dot{\phi}_{\mu} \approx \lambda_{\mu} \phi_{\mu} + \beta^2 n^{5/2} f(t)$.
It follows $|\phi_{\mu}|^2 \sim \beta^4 n^5 t$,
and the rate $D = 1/T \sim \beta^4 n^4$. The diffusion equation $m_2 \sim D t$ yields
\begin{equation}
m_2 \sim \beta^{4/3} t^{\alpha}\;,\; \alpha = 1/3\;.
\label{subdif}
\end{equation}
The predicted exponent is close to the numerically observed one.

In order to clarify the nature of resonant mode pairs, we studied statistical properties
of resonant pairs for $W=4$ with $R_{\nu,\mu_0} < x_c$. Here $x_c \sim \beta n$ is a cutoff value, which
decreases the more the packet spreads (in our simulations we reach values $x_c \sim 0.01$).
The corresponding distributions of the overlap integral $|I_{\mu_0,\nu,\nu,\nu}|$ appear to
be invariant, with an average value $\overline{I} \approx 0.05$. At the same time,
the distributions of the frequency spacing $\delta = |\lambda_{\nu}-\lambda_{\mu_0}|$ favour smaller
values the smaller  $x_c$ is, with an average spacing $\overline{\delta}  = 0.0026$ ($x_c=0.1$) and 
$\overline{\delta}  = 0.00026$ ($x_c=0.01$).
Both NMs from a resonant pair with small $R_{\nu,\mu_0}$ have a multipeak, or hybrid, structure
(inset in Fig.\ref{fig3} right plot). We evaluated the probability density $\mathcal{P}(d)$ of
the peak distance $d$, estimated with twice the square root of the second moment of the distribution
$z_l=A_{\nu,l}^2A_{\mu_0,l}^2$ (Fig.\ref{fig3} right plot). 
Its average increases from $\overline{d}=24.8$ for $x_c=0.1$
to $\overline{d} = 32.3$ for $x_c=0.01$. 
With the help of semiclassical tunneling rate estimates \cite{LLIII}
the level splitting of such a hybrid pair can be estimated
to be $ \delta \sim A_{\nu,l_0}^2$ where $l_0$ marks the midpoint between the peaks.
This leads to a splitting of the order of $\delta \sim {\rm e}^{-d/\xi}$. Therefore,
an increase of the packet size goes along with a much weaker increase of the distance $d$,
in accord with our explanation from above. 

Finally we consider the process of resonant excitation of an exterior mode by
a mode from the packet. The number of packet modes in  
a layer of the width of the localization volume at the edge, which are resonant with a
cold exterior mode,
will be proportional to $\beta n$. After long enough spreading $\beta n \ll 1$.
On average there will be no mode inside the packet, which could efficiently
resonate with an exterior mode. Therefore, resonant growth can be excluded.

The subdiffusive spreading is universal, i.e. the exponent $\alpha$ is independent of $\beta$
and $W$, which are only affecting the prefactor in (\ref{subdif}). 
Excluding selftrapping, any nonzero nonlinearity strength $\beta$ 
will completely delocalize the wavepacket and destroy Anderson localization.
The exponent $\alpha$ is determined solely by the degree of nonlinearity,
which defines the type of overlap integral to be considered in (\ref{PERT1}),
and by the stiffness of the spectrum $\{ \lambda_{\nu} \}$.
We performed fittings by analyzing 20 runs in regime ii) with different disorder
realizations. For each realization we fitted the exponent $\alpha$, and
then averaged over all computational measurements. 
We find $\alpha = 0.33 \pm 0.02$ for DNLS,
and $\alpha = 0.33 \pm 0.05$ for KG.
Therefore, the predicted universal exponent $\alpha=1/3$ explains all available data.

Using the above approach, we estimate the growth of the second moment of
a wavepacket which is excited on top of a nonzero norm density background $n_0$. 
Assuming the wavepacket having norm density $(n+n_0)$
we find the rate $D \sim \beta^4 (n+n_0)^4$ and therefore the second moment of the 
wavepacket $m_2 \sim 1/n^2 \sim \beta^4 (n+n_0)^4 t$. It follows that as long as
$n \gg n_0$ holds, the wavepacket spreads subdiffusively 
$m_2  \sim \beta^{4/3}  t^{1/3}$. But once the wavepacket decays such that
$n \sim n_0$,  normal diffusion 
$m_2 \sim \beta^4 n_0^4 t$ sets in, in accord with \cite{abjl08}. 
The crossover time scales to infinity as 
$n_0$ is approaching zero. 

Let us generalize our results to $d$-dimensional lattices with
nonlinearity order $\sigma > 0$:
\begin{equation}
i\dot{\psi_{\boldsymbol{l}}}= \epsilon_{\boldsymbol{l}} \psi_{\boldsymbol{l}}
+\beta |\psi_{\boldsymbol{l}}|^{\sigma}\psi_{\boldsymbol{l}}
-\sum\limits_{\boldsymbol{m}\in
D(\boldsymbol{l})}\psi_{\boldsymbol{m}}\;.
\label{RDNLS-EOM}
\end{equation}
Here $\boldsymbol{l}$ denotes a $d$-dimensional lattice vector with
integer components, and $\boldsymbol{m}\in
D(\boldsymbol{l})$ defines its set of nearest neighbour lattice sites.
We assume that (a) all NMs are spatially localized (which can be obtained for strong
enough disorder $W$), and (b) the property $\mathcal{W}(x \rightarrow 0) \rightarrow const \neq 0$
holds. A wavepacket with average norm $n$ per excited mode has a second moment
$m_2 \sim 1/n^{2/d}$. It follows \cite{deph}  
\begin{equation}
m_2 \sim \left(\beta^{4} t\right)^{\alpha}\;,\;\alpha=\frac{1}{1+d\sigma}\;.
\label{generalization}
\end{equation}
The exponent $\alpha$ is bounded from above by $\alpha_{max}=1$, which
is obtained for $\sigma \rightarrow 0$.
For the above studied two-body interaction $\sigma=2$ we predict
$\alpha(d=2)=1/5$ and $\alpha(d=3)=1/7$.

It is a challenging task to determine the bounds of the domains of validity of (\ref{generalization}).
They will be reached when $\mathcal{W}(x \rightarrow 0) \rightarrow 0$.
Is further growth prohibited then?
We think not, because trapping an excitation in a finite volume must generically
lead to equipartition on finite times due to nonintegrability.
That induces a finite (whatever weak) chaotic component in the 
dynamics, which will heat the cold exterior. Extending the above resonance scenario,
higher order resonances are expected to persist and to yield further (though slower) spreading.
%
%
\\
\\
We thank S. Komineas for very intensive help, and B. L. Altshuler, S. Aubry, L. Schulman and W.-M. Wang
for useful discussions.


\begin{thebibliography}{99}
\bibitem{equiv}
For small amplitudes the equations of motion of the KG chain can be
approximately mapped onto a corresponding DNLS model
(see \cite{DB} and references therein). In our notations,  the norm density
of the DNLS model corresponds to the normalized energy density
of the KG model and $\beta\approx 3WE$.
\bibitem{numerics}
The system size was typically $N=1000$ or larger.
Excitations did not reach the boundaries during the integration time, and results are 
unchanged when further increasing $N$. Symplectic integrators were used for both models
to have boundness of the errors. We also varied time step sizes to ensure that 
results are not influenced by roundoff errors.
\bibitem{comment1}
Argumentations in \cite{PS08} yield an incorrect exponent $\alpha=2/5$, because
the heating rate of the cold exterior was assumed to be equal to the packet diffusion
rate, which it is not.
\bibitem{comment2}
Each 100 time units on average 50\% of the NMs were randomly chosen, and their phases
were shifted by $\pi$ (DNLS). For the KG case we changed the signs of the corresponding
NM momenta.
\end{thebibliography}

\end{document}